\documentclass[aps,prl,twocolumn,showpacs,superscriptaddress]{revtex4-1}
\usepackage{color}
\usepackage{graphicx}
\usepackage{amsmath}
\usepackage{amsfonts}
\usepackage{amssymb}
\newcommand{\ket}[1]{|#1\rangle}
\newcommand{\bra}[1]{\langle #1|}

\begin{document}
\today
\title{Unifying Geometric Entanglement and Geometric Phase in a Quantum Phase Transition}
\author{Vahid Azimi Mousolou\footnote{Electronic address: vahid.mousolou@lnu.se}}
\affiliation{Department of Physics and Electrical Engineering, Linnaeus University,
391 82 Kalmar, Sweden}
\author{Carlo M. Canali\footnote{Electronic address: carlo.canali@lnu.se}}
\affiliation{Department of Physics and Electrical Engineering, Linnaeus University,
 391 82 Kalmar, Sweden}
\author{Erik Sj\"oqvist\footnote{Electronic address: erik.sjoqvist@kvac.uu.se}}
\affiliation{Department of Quantum Chemistry, Uppsala University, Box 518,
Se-751 20 Uppsala, Sweden}
\affiliation{Centre for Quantum Technologies, National University of Singapore,
3 Science Drive 2, 117543 Singapore, Singapore}
\begin{abstract}
Geometric measure of entanglement and geometric phase have recently been used to 
analyze quantum phase transition in the XY spin chain. We unify these two approaches by 
showing that the geometric entanglement and the geometric phase are respectively 
the real and imaginary parts of a complex-valued geometric entanglement, which can  
be investigated in typical quantum interferometry experiments. We argue that the singular 
behavior of the complex-value geometric entanglement at a quantum critical point is 
a characteristic of any quantum phase transition, by showing that the underlying mechanism 
is the occurrence of level crossings associated with the underlying Hamiltonian.
\end{abstract}
\maketitle
Quantum phase transitions (QPTs) are qualitative changes in the properties of many-body 
systems driven by quantum fluctuations at zero temperature. A key feature to understand 
the nature of QPTs is the quantum correlations between the system degrees of freedom, 
which at the critical point bring about an intersection of the ground-state and excited-state energy levels. 
While QPTs have traditionally been characterized by appropriate correlation functions \cite{sachdev99,henkel99}, many researchers have recently addressed this problem from 
a quantum information perspective. One idea along this line is to use measures of entanglement 
to characterize QPTs. Indeed, there has been a great deal of work analyzing properties of 
quantum entanglement in many-body systems undergoing QPT \cite{osterloh02,wei05,orus08}. 
Another idea is based on the fact that non-trivial geometric phases (GPs) are strongly 
affected by quantum and classical correlations residing in the many-body quantum states 
\cite{williamson07}, and especially by the level degeneracies associated with QPTs in such 
systems. Therefore GPs can also be used to detect the presence of a QPT. Relations between 
GP and QPTs have been established theoretically \cite{zhu06, carollo05, gp-qpt} and 
experimentally in NMR \cite{peng10}.

In this paper, we examine the  geometric measure of entanglement (GE) 
\cite{wei05,shimony95,wei03} and GPs in the vicinity of QPT in a XY spin-chain. We determine a 
relation between these geometric objects, allowing us to identify them as the real and 
imaginary part of a single measurable quantity, which we call the complex-valued GE. 
We demonstrate that the complex GE is accessible in interferometry experiments. 
The established relation is valid for a general quantum many-body system. At a 
quantum critical point both real and imaginary parts of the complex-valued GE display 
the same singular behavior, which in turns is closely associated with the singular behavior 
of the quantum geometric tensor triggered by a level degeneracy. This provides a universal 
approach to the study of quantum critical phenomena. 
    
Consider the one-dimensional XY model system with $N$ sites in a transverse magnetic 
field. The corresponding Hamiltonian reads
\begin{eqnarray}
H(r,h) =  - \sum_{j=1}^N \left( \frac{1+r}{2} \sigma^x_j \sigma^x_{j+1} \right.
\left. + \frac{1-r}{2}  \sigma^y_j \sigma^y_{j+1} +
h \sigma^z_{j}\right)
\nonumber \\
\label{xymodel}
\end{eqnarray}
with periodic boundary condition such that the first and $(N+1)$th sites are identified. 
Here, $r$ is the anisotropy parameter, $h$ is the magnetic field strength, and $\sigma_{j}^k$, 
$k=x,y,z$, are the standard Pauli operators of the $j$th spin. In the thermodynamic limit 
($N\rightarrow\infty$), the system described by Eq.~(\ref{xymodel}) undergoes a QPT
at $h=h_c= 1$. The full phase diagram can be found in Refs. \cite{bm71,henkel99}. In 
the anisotropy range $0<r\leq1$ the system belongs to the Ising universality class. The 
isotropic case $r = 0$ corresponds to the XX model, which belongs to the 
Berezinsky-Kosterlitz-Thouless universality class. The standard procedure to solve 
the eigenvalue problem of $H(r,h)$ is to convert the spin operators into fermionic operators, 
using successively Jordan-Wigner, Fourier, and Bogoliubov transformations \cite{lieb61}. 
The ground state in the even fermion number sector can be expressed in terms 
of fermionic modes $c_j$ as
\begin{eqnarray}
\ket{\psi (r,h)} & = & \prod_{m=0}^{m<\frac{N-1}{2}} \big[ \cos \theta_m (r,h) 
\nonumber \\
 & & +i\sin\theta_m (r,h) c_m^{\dagger} 
c_{N-m-1}^{\dagger} \big] 
\ket{\uparrow...\uparrow} 
\label{groundstate}
\end{eqnarray}
with $\tan 2\theta_m (r,h) = r\sin\frac{\pi(2m+1)}{N}/(h-\cos\frac{\pi(2m+1)}{N})$ 
\cite{wei05}.
For $0< r\le 1$ the ground-state in the thermodynamic limit is doubly degenerate in the 
ferromagnetic regime $h< 1$ and singly degenerate for $h > 1$, with the critical value
for $h_{c}= 1$ being a point of conical intersection.

Entanglement of the ground state $\ket{\psi(r,h)}$ can be measured by approximating it 
by  the closest pure product state \cite{wei03}. This state is given by the maximal overlap
\begin{eqnarray}
\Lambda_{\max}(r,h)=\max_{\Phi}\left|\left\langle \Phi|\psi(r,h)\right\rangle\right|,
\end{eqnarray}
over all pure product states $\Phi$. The resulting number $\Lambda_{\max} (r,h)$ is the 
entanglement eigenvalue of the XY ground state \cite{wei03}. GE of the ground state is quantified via \cite{wei05}
\begin{eqnarray}
E_{\log_2} [\psi(r,h)] = -\log_2\Lambda_{\max}^2(r,h),
\end{eqnarray}
which in the thermodynamic limit is characterized by the geometric entanglement (GE) density 
\begin{eqnarray}
\varepsilon(r,h) = 
\lim_{N\rightarrow\infty}\varepsilon_N(r,h) ,
\label{entanglement-density}
\end{eqnarray}
where $\varepsilon_N(r,h) = E_{\log_2} [\psi(r,h)] /N$ is the entanglement per site. 
Using the translational symmetry of $\ket{\psi (r,h)}$, the closest pure 
product state for each value of $r$ and $h$ takes the form \cite{wei05} 
\begin{eqnarray}
 \ket{\Phi(\xi)} & = & \otimes_{j=1}^N 
\left[ \cos \left( \frac{\xi}{2} \right) \ket{\uparrow}_j + 
\sin \left( \frac{\xi}{2} \right) \ket{\downarrow}_j \right],
\label{separable-state}
\end{eqnarray}
where the parameter $\xi$ is chosen so as to maximize the overlap, i.e., $\Lambda_{\max} (r,h) = 
\max_{\xi}\left|\left\langle \Phi (\xi)|\psi(r,h)\right\rangle\right| \equiv \left|\left\langle 
\Phi (\xi_{\max})|\psi(r,h)\right\rangle\right|$. By using
Eq.~(\ref{entanglement-density}), one obtains the GE density \cite{wei05} 
\begin{eqnarray}
\varepsilon(r,h) & = & -\frac{2}{\ln 2}\max_{\xi}\int_{0}^{1/2}
d\mu\ln[\cos\theta(\mu,r,h)\cos^2(\xi/2) 
\nonumber\\
 & & + \sin\theta(\mu,r,h)\sin^2(\xi/2)\cot\pi\mu],
\end{eqnarray}
with $\tan2 \theta(\mu,r,h) = r\sin2 \pi\mu/(h-\cos2 \pi\mu)$.

\begin{figure}[h]
\centering
\includegraphics[width=55mm,height=37mm]{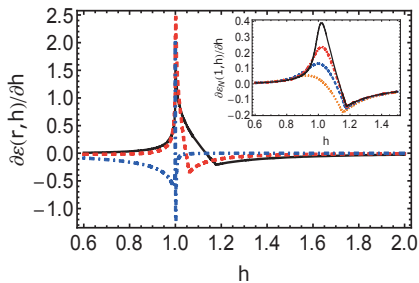}
\caption{(Color online) Derivative of the ground-state GE density $\varepsilon (r,h)$
of the XY model as function of the magnetic field $h$ in the thermodynamic 
limit. Solid curve: Ising limit ($r=1$); dashed curve: anisotropic ($r=0.5$) XY model; dot-dashed 
curve: anisotropic ($r=0.05$) XY model. The inset corresponds to the Ising limit for different finite 
lattice size $N=11, 18, 31, 64$.}
\label{fig:dGE}
\end{figure}

It has been shown \cite{wei05} that features of quantum criticality, such as universality, critical 
exponents, and scaling are captured by the singular behavior of the GE of the XY ground state. 
The critical exponents for different universality classes have been found by scaling analysis of 
divergences at the singular points of the GE density. The singular behavior of the 
GE density at the critical point $h_c=1$ is shown in Fig. \ref{fig:dGE}, where 
$\partial \varepsilon_N (r,h) / \partial h$ has been plotted as function of magnetic field $h$ 
for different values of $r$ in the thermodynamic limit. In the inset we plot the same quantity 
for Ising model for finite values of $N$. The cusp at $h_s> 1$ for $0< r \le 1$, coalescing 
with the singularity at $h_{c}=1$ as $r\to 0$, is an interesting feature related to properties 
of the closest product state $\Phi(\xi_{\max})$. 
For each $0 \le r\le 1$ there is a finite value of $h = h_s$ for which $\xi_{\max}$ vanishes and  
$\Phi(\xi_{\max})$ is polarized in the $z$-direction \cite{wei05}. This state remains the closest 
pure product state for all $h > h_s$, causing a discontinuity jump in the second derivative 
of $\varepsilon(r,h)$ at $h = h_s$.

\begin{figure}[h]
\centering
\includegraphics[width=50mm,height=35mm]{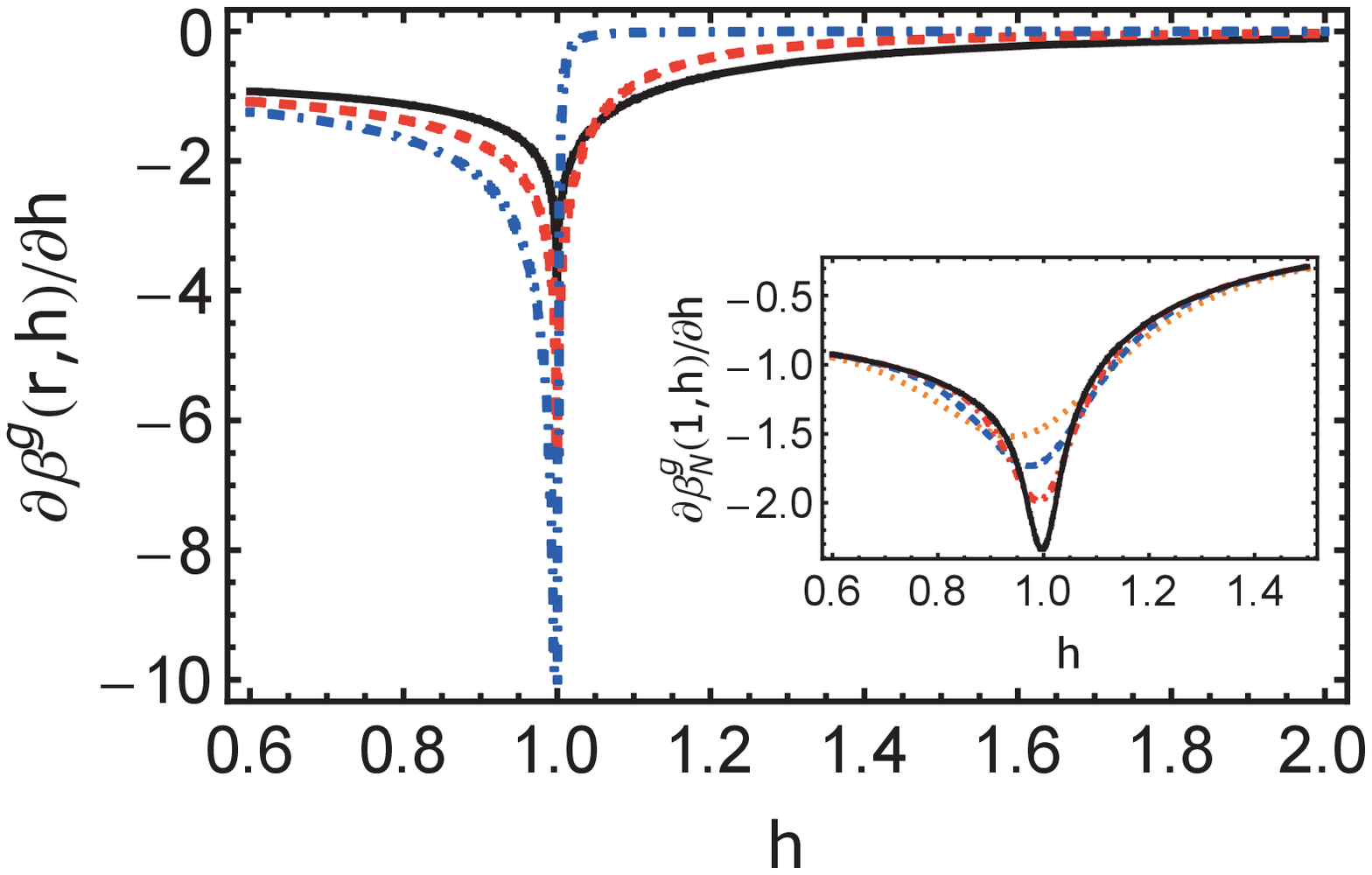}
\includegraphics[width=50mm,height=35mm]{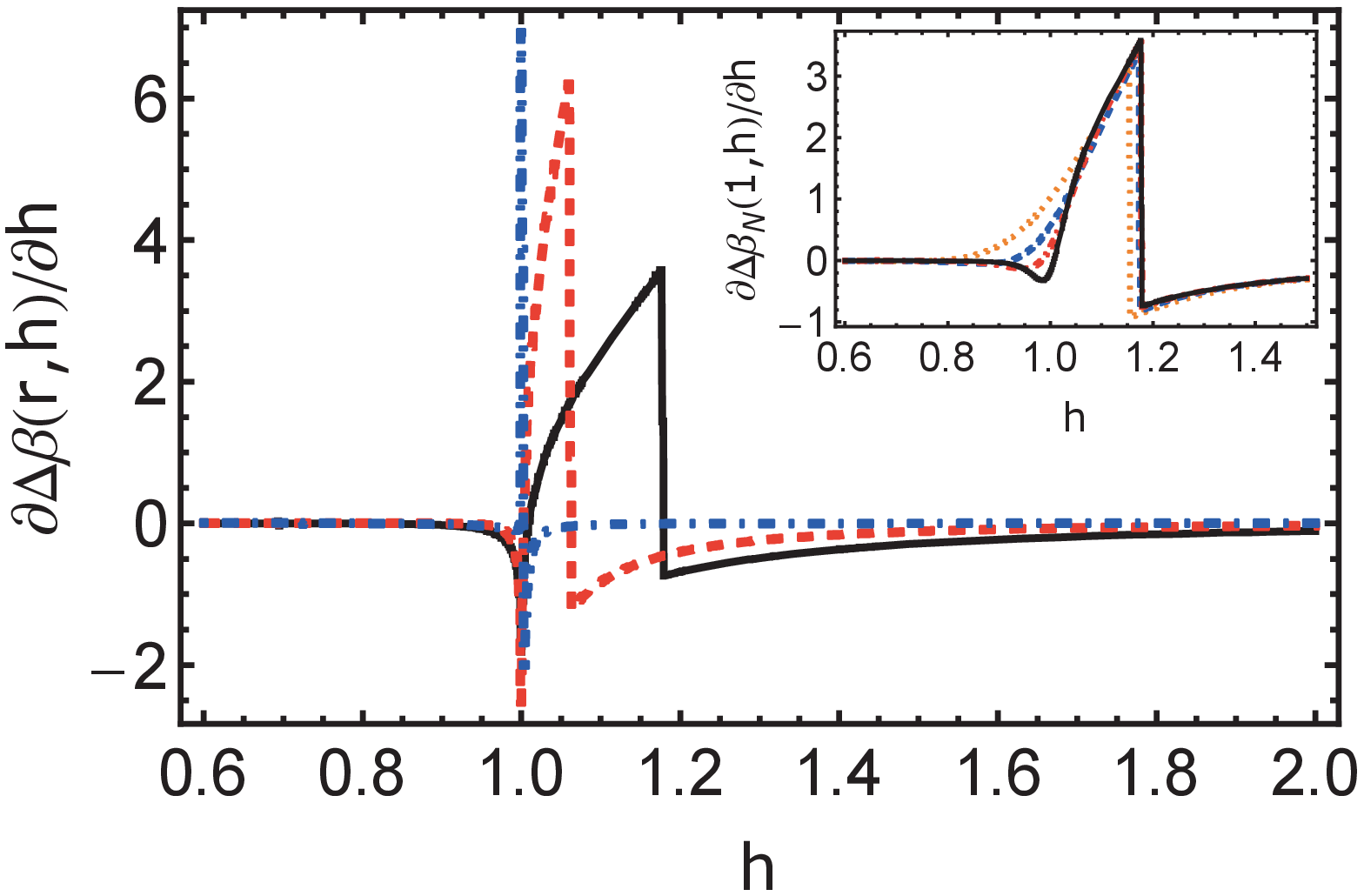}
\caption{(Color online) Top panel: Derivative of the GP density, GP per particle, corresponding 
to cyclic evolution of the XY ground state given in Eq. (\ref{g-loop}), as function of 
magnetic field $h$ in the thermodynamic limit. Lower panel: $h$ derivative of difference between 
GP densities corresponding to the evolutions given in Eqs. (\ref{g-loop}) and (\ref{p-loop}), plotted 
as function of magnetic field $h$ in the thermodynamic limit. Solid curve: Ising limit ($r=1$); dashed 
curve: anisotropic ($r=0.5$) XY model; dot-dashed: anisotropic ($r=0.05$) XY model. The insets 
correspond to Ising limit for different finite lattice size $N=11, 18, 31, 64$.}
\label{fig:dGP}
\end{figure}

Recently, a generic connection between GPs and QPTs has been identified \cite{zhu06}. A 
scaling analysis of GP of the XY ground state reveals similar information about quantum 
criticality as the GE density. Top panel in Fig. \ref{fig:dGP} shows the behavior 
of the GP per site $\beta_N^g (r,h)$ of the XY ground state, accumulated by adiabatically 
rotating each spin around the $z$ axis via
\begin{eqnarray}
 & C_g^{(r,h)} : [0,\pi] \ni \phi \rightarrow
\nonumber \\
 & U_{{\textrm{tot}}}(\phi) \ket{\psi(r,h)} \bra{\psi(r,h)}
U_{{\textrm{tot}}}^{\dagger}(\phi),
\label{g-loop}
\end{eqnarray}
where $U_{{\textrm{tot}}}(\phi)=\otimes_{j=1}^NU_{j}(\phi)  = \otimes_{j=1}^N 
\exp(\frac{i\sigma^z_{j}\phi}{2})$, for different lattice size $N$. $\beta^g (r,h) = 
\lim_{N\rightarrow \infty} \beta_N^g (r,h)$ is the GP density, i.e., the GP per site 
in the thermodynamic limit.

The lower panel of Fig. \ref{fig:dGP} displays the behavior of the difference between the 
GPs per site associated with $C_g^{(r,h)}$ and the evolution
\begin{eqnarray}
& C^{(r,h)}_p : [0,\pi] \ni \phi \rightarrow
\nonumber \\
& U_{{\textrm{tot}}}(\phi) \ket{\Phi(r,h)} \bra{\Phi(r,h)}
U_{{\textrm{tot}}}^{\dagger}(\phi)
\label{p-loop}
\end{eqnarray}
of the closest pure product state $\ket{\Phi (r,h)} \equiv \ket{\Phi (\xi_{\textrm{max}})}$. 
$C^{(r,h)}_p$ is a path of pure product states along which each instant state is the closest 
pure product state of the corresponding instant ground state along the path $C_g^{(r,h)}$. 
The entanglement is fixed along this pair of paths. The associated GP difference is in this 
sense related to the entanglement residing in the initial ground state $\ket{\psi(r,h)}$.

In the thermodynamic limit, we obtain the GP densities 
\begin{eqnarray}
\beta^g (r,h) & = & \lim_{N\rightarrow \infty} \beta_N^g (r,h) 
\nonumber \\ 
 & = &  2\pi\int_0^{1/2}\sin^2\theta(\mu,r,h)d\mu,
\nonumber\\
\beta^p (r,h) & = & \lim_{N\rightarrow \infty} \beta_N^p (r,h) = 
2\pi \sin^2 (\xi_{\max}/2)
\label{Berry-phases}
\end{eqnarray}  
corresponding to $C_g^{(r,h)}$ and $C_p^{(r,h)}$, respectively. 

The GP density difference is
\begin{equation}
\Delta\beta(r,h) \equiv \beta^g (r,h)-\beta^p (r,h).
\end{equation}
For $0 < r\le1$, $\Delta\beta(r,h)$ displays the same
(Ising-universality class) critical behavior at $h= h_c$ as
$\varepsilon(r,h)$. 
Interestingly, even the cusp in $\varepsilon (r,h)$ at $h_s>h_c$ is captured by 
$\Delta \beta (r,h)$ (now in the form of a discontinuous jump  in the first derivative 
of $\Delta \beta (r,h))$, while the ground state GP density $\beta^g (r,h)$ is smooth 
on this $h$ interval.  

For the XX model ($r=0$), we observe that $\Delta\beta(0,h)=0$ and therefore 
$\frac{\partial}{\partial h} \Delta\beta(0,h)=0$ for the specific evolution operator 
$U_{\textrm{tot}} (\phi)$ \cite{note}. On the other hand, as indicated in the lower panel of 
Fig. \ref{fig:dGP}, $\frac{\partial}{\partial h} \Delta\beta(r,h)$ is singular at $h=h_c=1$ 
when $r\ne 0$. Thus,  
\begin{eqnarray}
\lim_{r\rightarrow 0} \left. \frac{\partial\Delta\beta(r,h)}{\partial h}\right|_{h=h_{c}} \neq 
\left. \frac{d\Delta\beta(0,h)}{dh}\right|_{h=h_{c}} ,
\label{GP-Diff-xx}
\end{eqnarray}
which shows that the XX criticality is detected by non-smooth contractible phase difference. 
Arguments along this line have been used in Refs. \cite{carollo05,hamma06} to locate the 
criticality of XX model by using GP. From the analysis of Ref. \cite{wei05}, we see that also 
$\partial\varepsilon(r,h)/\partial r$ scales differently for $r\ne0$ and $r=0$ near the 
critical point. Just as in the case of GE density, the nature of Eq. (\ref{GP-Diff-xx}) 
identifies the different universality classes of XY spin chain.  

We now give a unifying operational interpretation of GP and GE for the XY system, in the context 
of interferometry. Let $\ket{0},\ket{1}$ span the state space of an auxiliary qubit, playing the role 
of the two interferometer arms. Prepare the initial state $\ket{\Psi_i}=\frac{1}{\sqrt{2}} 
(\ket{0}+\ket{1}) \ket{\psi (r,h)}$. In the $\ket{1}$ arm, project onto the product state 
$\ket{\Phi (\xi)}$ given in Eq. (\ref{separable-state}). The two spin chain states $\ket{\psi (r,h)}$ 
and $\ket{\Phi (\xi)}$ are exposed to the parallel transporting unitary operators 
$U_{\psi}^{\parallel} (\phi) = \otimes_{j=1}^N e^{\frac{i}{2} \phi (\sigma_j^z - \bra{\psi (r,h)} 
\sigma_j^z \ket{\psi (r,h)})}$ and $U_{\Phi}^{\parallel} (\phi) = \otimes_{j=1}^N e^{\frac{i}{2} 
\phi (\sigma_j^z - \bra{\Phi (\xi)} \sigma_j^z \ket{\Phi (\xi)})}$, respectively, where 
$\phi \in [0,\pi]$. Note that $U_{\psi}^{\parallel} (\phi)$ and $U_{\Phi}^{\parallel} (\phi)$ 
is the same unitary up to an overall phase factor, which assures that 
$|\bra{\Phi (\xi)}U^{\parallel\dagger}_{\Phi} U_{\Psi}^{\parallel} \ket{\Psi (r,h)}|$ for given $r$ 
and $h$ is a function of $\xi$ only. The resulting state $U_{\Phi}^{\parallel} (\pi) 
\ket{\Phi (\xi)}$ 
in the $\ket{1}$ arm is parallel transported along the shortest geodesic back to $\ket{\Phi (\xi)}$, 
while a U(1) shift $e^{-if}$ is applied to the $\ket{0}$ arm. Finally, the two arms are brought back 
to interfere, resulting in the final state 
\begin{eqnarray} 
\ket{\Psi_f} & = & \frac{1}{2} \ket{0} \left( e^{i(\varphi^g - f)} \ket{\psi (r,h)} \right. 
\nonumber \\ 
 & & \left. + e^{i\varphi^p} \ket{\Phi (\xi)} \langle \Phi (\xi) \ket{\psi (r,h)} \right) + \ldots , 
\end{eqnarray}
with $\varphi^g$ and $\varphi^p$ being the GPs associated with the evolution of 
$\ket{\psi (r,h)}$ and $\ket{\Phi (\xi)}$, respectively. The intensity in the $\ket{0}$ 
arm becomes 
\begin{eqnarray} 
\mathcal{I}_0 & = & \frac{1}{4} \left( 1+ \left| \langle \Phi (\xi) \ket{\psi (r,h)} \right|^2 \right) 
\nonumber \\ 
 & & + \frac{1}{2} {\textrm{Re}} \left( A(r,h;\xi) e^{-if} \right) ,
\end{eqnarray}
where 
\begin{eqnarray}
A(r,h;\xi) & = & \bra{\psi (r,h)} e^{-i\varphi^p} \ket{\Phi (\xi)} 
\bra{\Phi (\xi)} e^{i\varphi^g} \ket{\psi (r,h)} 
\nonumber \\ 
 & = & \left| \langle \Phi (\xi) \ket{\psi (r,h)} \right|^2 e^{i\Delta\varphi} . 
\end{eqnarray}
Here, $|A(r,h;\xi)| = \left| \langle \Phi (\xi) \ket{\psi (r,h)} \right|^2$ and $\arg A(r,h;\xi) = 
\Delta \varphi$ characterize the visibility and phase shift of the interference fringes obtained 
by varying $f$. Note that $A(r,h;\xi)$ is unchanged under any local gauge transformation and 
reparametrization of the paths. The entanglement eigenvalue $\Lambda_{\max}(r,h)$ and the GP 
difference $\Delta \varphi (r,h)$ associated with $C_g^{(r,h)}$ and $C_p^{(r,h)}$ can be read-out 
from the interference fringes by tuning the single parameter $\xi$ to the value $\xi_{\max}$ 
giving the maximal visibility. 

Following the procedure leading to GE, we introduce a complex-valued GE being defined as 
the following extensive entanglement sensitive quantity
\begin{eqnarray}
E^c_{\log_2} [\psi(r,h)] & = & -\log_2 A(r,h;\xi_{\max})
\nonumber\\
 & = & E_{\log_2} [\psi(r,h)] - i\frac{\Delta\varphi(r,h)}{\ln 2} . 
\label{complex-valued-GE-xy}
\end{eqnarray}
The properties of the quantum critical point is characterized by both the real and imaginary 
parts of the complex-valued GE density 
\begin{eqnarray}
\varepsilon_c(r,h) & = & \lim_{N\rightarrow\infty}\frac{1}{N}E^c_{\log_2} [\psi(r,h)]
\nonumber\\
 &=&\varepsilon(r,h)-i\frac{\Delta\beta(r,h)}{\ln 2}.
\label{complex-valued-ED-xy}
\end{eqnarray}
The GE and GP densities are thus two different sides of the same coin: the 
complex-valued GE density $\varepsilon_c (r,h)$. 

\begin{figure}[h]
\centering
\includegraphics[width=45mm,height=25mm]{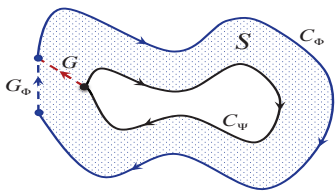}
\caption{Paths $C_{\Psi}$ and $C_{\Phi}$ of many-body state $\Psi$ and its closest product state 
$\Phi$, respectively. $G$ is a geodesic in state space projecting $\Psi$ onto $\Phi$.}
\label{fig:paths}
\end{figure}

Let us now generalize the above ideas to a generic $N$-body system. Assume that the 
system is prepared in the state $\ket{\Psi}$ and thereafter evolves around a loop $C_{\Psi}$ 
generated by a one-parameter family of local unitary operators $U(t)=\otimes_j^N U_j(t)$ with 
$U_j(t)$ acting on subsystem $j$. The closest product state $\ket{\Phi}$  
evolves under the same $U(t)$. Note that the resulting $C_{\Phi}$ is in general an open path.  
Let $\Delta\varphi(\Psi) = \varphi(\Psi) - \varphi(\Phi)$ be the GP difference associated 
with paths $C_{\Psi}$ and $C_{\Phi}$. By following the above interferometer scheme, 
the interference fringes are characterized by 
\begin{eqnarray} 
A(\Psi) = \Lambda_{\max}^2(\Psi) e^{i\Delta\varphi(\Psi)} 
\label{interference-general}
\end{eqnarray} 
with $\ket{\Phi}$ chosen to maximize the visibility. Similar to the XY-case discussed above, 
$A(\Psi)$ can be used to define complex-valued GE of the $N$-body system prepared in 
$\Psi$. Furthermore, $A(\Psi)$ can be interpreted in terms of the Hermitian metric  
$\mathcal{T}$ on the projective state manifold $\mathcal{P}(\mathcal{H})$ \cite{provost80}. This metric tensor is defined as 
$\mathcal{T}(v,u) = \bra{v}(1-\ket{\psi}\bra{\psi})\ket{u}$ for tangent vectors $u$ and $v$ at a 
state $\ket{\psi}\bra{\psi}\in\mathcal{P}(\mathcal{H})$. We can write $\mathcal{T}=\mathcal{G}-i\mathcal{F}$, where $\mathcal{G}$ 
is the Riemannian metric providing the Fubini-Study measure of distance in $\mathcal{P}(\mathcal{H})$ 
and $\mathcal{F}$ is the symplectic curvature 2-form responsible for the GP. Let $ds^2=\mathcal{G}(v,v)$ for each tangent vector $v$ to $\mathcal{P}(\mathcal{H})$ denote the square of the line element associated with metric $\mathcal{G}$. Then, we have  
\begin{eqnarray}
\Lambda_{\max}(\Psi)&=&1-\frac{1}{2}\left[\int_{G} ds\right]^2, 
\nonumber\\
\Delta\varphi(\Psi)&=&\int_{S}\mathcal{F},
\label{GE-GP}
\end{eqnarray}
where $G$ is a geodesic in state space connecting $\Psi$ and $\Phi$, and $S$ is the 
oriented surface with boundary $\partial S =  C_{\Psi} \ast G \ast G_{\Phi}^{-1} \ast C_{\Phi}^{-1} 
\ast G^{-1}$. The inverse denotes the reverse direction along the corresponding path (see Fig. \ref{fig:paths}). Equation (\ref{GE-GP}) provides a physical realization of the Hermitian 
metric $\mathcal{T}$ in terms of the complex-valued GE, which is accessible in an interferometer.

A general relation between QPTs and the geometrical objects $ds$ and $\mathcal{F}$ have been 
pointed out in Refs. \cite{venuti07, zhu06}, where the external parameter $\lambda\in\mathcal{M}$ 
was incorporated into the system. This relation can be understood basically by pulling back the 
Hermitian metric $\mathcal{T}$ on $\mathcal{M}$ via the map $\Psi_0\ :\ 
\mathcal{M}\ni\lambda\longrightarrow \ket{\psi_0(\lambda)}\bra{\psi_0(\lambda)}\in\mathcal{P}
(\mathcal{H})$, where $\ket{\psi_0(\lambda)}$ is the unique ground state of the corresponding 
Hamiltonian $H(\lambda)$. For a coordinate system on $\mathcal{M}$, the components of the 
pull-back tensor $T=\Psi_0^*\mathcal{T}$ are 
\begin{eqnarray}
T_{\mu\nu}=\sum_{n\neq0}\frac{\bra{\psi_0(\lambda)}\partial_\mu H\ket{\psi_n(\lambda)}\bra{\psi_n(\lambda)}\partial_\nu H\ket{\psi_0(\lambda)}}{[E_0(\lambda)-E_n(\lambda)]^2},\hspace{3mm}
\label{QGT}
\end{eqnarray}
where $\psi_n(\lambda)$ are eigenstates of the system and the indices $\mu, \nu=1, \ldots , \text{dim}\mathcal{M}$ are labeling the coordinates of $\mathcal{M}$. The tensor $T$ is known 
as quantum geometric tensor, whose imaginary part is the Berry curvature on $\mathcal{M}$ 
\cite{berry84} and real part provides an approximation of the fidelity of two ground states 
associated to neighboring points on $\mathcal{M}$ \cite{venuti07}. Unlike the physical interpretation 
of $T$, the above interpretation of $\mathcal{T}$ as complex-valued GE is exact and does not 
depend on any approximation.

The energy denominators $E_n(\lambda)-E_0(\lambda)$ in Eq. (\ref{QGT}) show that in the 
thermodynamic limit, 
$\textrm{Re} T$ and $\textrm{Im} T$ will show a singular behavior at the QPT as a QPT occurs at level crossing or avoided level crossing. This 
singularity is then reflected in $\Lambda_{\max}(\psi_0(\lambda))$ and $\Delta\varphi(\psi_0(\lambda))$ through the expressions given in Eq. (\ref{GE-GP}). Therefore it can be captured by both GE and GP difference, which are respectively the real and imaginary parts of the complex-valued GE. This source of singularity has been brought up
in the Berry phase and fidelity approaches \cite{venuti07, zhu06}. Our analysis suggests that the critical behavior at QPT in the GE and thus in the complex-valued GE of a generic many-body ground state comes from the same source. 

In summary, we have established a unifying connection between the GP and the GE in a generic 
many-body system. They are respectively the real and imaginary part of a generalized complex-valued 
GE and they both become singular at a QPT. We have given an exact geometrical interpretation of 
the complex-valued GE in terms of the induced Hermitian metric on the projective Hilbert state 
space. Finally, we have proposed one common source for the critical behavior of the GE at any QPT.  
The approach presented here can be tested experimentally in an interferometry setup, 
where the geometric entanglement would then yield the visibility of the interference fringes, whereas
the geometric phase would describe the phase shifts. The resulting interference fringes would 
therefore display large fluctuations at the critical point. Test-bed experiment to demonstrate 
the relevance of the complex-valued geometric entanglement in translational symmetric few-qubit 
ground states would be within reach in, e.g., NMR systems. 

We wish to thank Paolo Zanardi for discussions and useful comments. 
C.M.C. and V.A.M were supported by Department of Physics and Electrical Engineering at 
Linnaeus University (Sweden) and by the National Research Foundation (VR). E.S. acknowledges 
support from the National Research Foundation and the Ministry of Education (Singapore).


\begin{thebibliography}{99}
\bibitem{note} Note that the GE density $\varepsilon(0,h)$ vanishes above the critical point  $h>h_{c}=1$, 
while it is positive below $h_{c}$ \cite{wei05}. So, $\ket{\psi(0,h)}$ and $\ket{\Phi(0,h)}$ coincide only above $h_{c}$; below $h_c$ these two states differ and may therefore pick up different GPs under general local unitary evolutions that differ from $U_{\textrm{tot}}(\phi)$. Therefore $|\Delta\beta(0,h<h_{c})|\geq 0$ while $\Delta\beta(0,h>h_{c})= 0$ for a general local unitary evolution.
\end{thebibliography}
\end{document}